\documentclass[12pt]{article}
\usepackage{amssymb,graphicx}

\makeatletter
\renewcommand\section{\@startsection {section}{1}{\z@}%
                                 {-3.5ex \@plus -1ex \@minus -.2ex}
                                   {2.3ex \@plus.2ex}%
                                   {\normalfont\large\bfseries}}
\renewcommand\subsection{\@startsection{subsection}{2}{\z@}%
                                   {-3.25ex\@plus -1ex \@minus -.2ex}%
                                     {1.5ex \@plus .2ex}%
                                     {\normalfont\bfseries}}
\renewcommand\subsubsection{\@startsection{subsubsection}{3}{\z@}%
                                   {-3.25ex\@plus -1ex \@minus -.2ex}%
                                     {1.5ex \@plus .2ex}%
                                     {\normalfont\itshape}}
\makeatother

\newcommand{\Letter}{
    \setlength{\textwidth}{7in}
    \setlength{\textheight}{9.5in}
    \hoffset=-0.75in
    \voffset=-1.15in }

\Letter

\newcommand{\gsim}{ \lower .75ex \hbox{$\sim$} \llap{\raise .27ex \hbox{$>$}} }
\newcommand{\lsim}{ \lower .75ex \hbox{$\sim$} \llap{\raise .27ex \hbox{$<$}} }

\begin{document}

\begin{titlepage}
\setcounter{page}{1} \baselineskip=15.5pt \thispagestyle{empty}
\begin{flushright}
\parbox[t]{2in}{
MAD-TH-08-04\\
}
\end{flushright}

\vfil

\begin{center}
{\LARGE Brane Inflation is Attractive}
\end{center}
\bigskip

\begin{center}
Bret Underwood$^\dagger$
\end{center}

\begin{center}
\textit{Department of Physics,
     University of Wisconsin,
     Madison, WI 53706, USA}
\end{center}

\bigskip \bigskip

\begin{center}
{\bf
Abstract}
\end{center}

We study the phase space of initial conditions for brane inflation, and find that
including the effects of the Dirac-Born-Infeld (DBI) kinetic term dramatically improves previous estimates
on the amount of fine tuning of initial conditions necessary for inflation, even for models dominated by slow roll.
Two effects turn out to be important for the phase space analysis: restrictions on the
total available phase space 
due to UV effects in
brane inflation,
and the extension of the inflationary attractor to the DBI inflationary regime.
We compare the amount of initial conditions fine tuning required for a brane inflation model
and its standard field theory counterpart 
and find that
brane inflation decreases the required tuning by several orders of magnitude.

\noindent 

\vfill
\hrulefill\hspace*{4in}

{\footnotesize $^{\dagger}$ Email:
\parbox[t]{7in}{bjunderwood@wisc.edu}}

\end{titlepage}

\newpage

\section{Introduction}
\label{sec:introduction}
\setcounter{equation}{0}

Inflation is a successful paradigm for generating a nearly scale-invariant, gaussian spectrum of primordial
density fluctuations \cite{Inflation,ChaoticInflation,NewInflation}, 
and many models are in excellent agreement with current data \cite{WMAP}.
Unfortunately, inflation is a paradigm without an underlying theory, since it is not known what role
the putative inflaton plays in the Standard Model or its extensions.

One potential problem arising from not knowing the origin of the inflaton field is how to specify
the initial conditions for the (pre)-inflationary era.  
From a 
conventional, classical, point of view, the
pre-inflationary era consists of a hot thermal bath, which may deviate significantly from isotropy and
homogeneity.  Here again the nature of the inflaton, in particular its coupling to the thermal bath
and self-coupling, is important for understanding the possible (in)homogenous initial conditions and their
subsequent evolution \cite{InhomogenAttractor}.  

The problem of inflationary initial conditions has been studied before \cite{InhomogenAttractor,SRAttractor}, with the result 
that large-field
inflationary models, such as chaotic inflation \cite{ChaoticInflation}, are typically stable to changes in initial conditions,
while small-field models, such as new inflation \cite{NewInflation}, are sensitive to initial conditions \cite{InhomogenAttractor}
(for more information
on the distinction between large- and small-field models, see \cite{LargeSmallModel}).  Hybrid inflation models \cite{HybridInflation}, 
which are driven
by a large vacuum energy density which decays through a separate tachyonic ``waterfall" field, also seem to be sensitive
to intial conditions because of their similarity to small field models.

Clearly, to make progress towards understanding the problem of inflationary initial conditions it would be helpful to know the
UV origin of the inflaton field and its behavior.  String theory is a compelling framework in which to build inflationary
models because of the ubiquity (at least at first glance) of weakly coupled scalar fields.  Furthermore, many simple models
of inflation suggest that the inflationary energy scale may be at or near the string scale, thus stringy physics may be
responsible both for the origin and initial conditions of the inflaton.

We will choose to focus on a class of inflationary models in string theory in which the inflaton is identified
as the position of a D-brane in an extra dimensional space, called brane inflation \cite{DvaliTye}.  
By now, many interesting
brane inflation models exist \cite{DD,KKLMMT,DelicateUniverse,Chasing,Uplifting,WrappedDBI,D3D7,Silverstein}.

One interesting common characteristic of brane inflation models is that their kinetic terms are of the Dirac-Born-Infeld
(DBI) form, which contains an infinite sum of higher derivative kinetic operators.  For a homogenous scalar field, the DBI
action imposes a speed limit on how fast the field can move which depends on the details of the geometry of the
extra dimensions.  For strongly warped spaces the speed limit can be severe, and can lead to a ``slowly rolling"
scalar field suitable for inflation even for very steep potentials \cite{Silverstein}.  Models in which the DBI kinetic term
plays an important role in generating inflation 
are called DBI inflation \cite{Silverstein}, and may be interesting observationally due to their potential
to probe details of the extra dimensions \cite{WarpedCMB}
and generate large primordial non-Gaussianity \cite{NonGauss} or gravity waves \cite{Bean}.

Within the paradigm of brane inflation, then, it seems natural to revisit the problem of initial conditions
for the inflaton field.  Some initial progress in this direction has been done \cite{InitialTuning} (see
also \cite{StochasticDBI} for recent work addressing the question of stochastic fluctuations
of the inflaton brane field), where it was found
that brane inflation models, falling into the class of hybrid and small-field inflation models, 
are sensitive to variations in (homogenous) initial
conditions, particularly to variations in the initial momentum of the inflaton field.  In this work, we will
take a step further, including the effects of the DBI kinetic term on the region of homogenous initial conditions
phase space which gives rise to sufficient inflation.  We will also see that the DBI kinetic term of
brane inflation models leads to an
extension of the standard slow roll inflationary attractor solution \cite{SRAttractor} to the DBI inflationary 
regime where the DBI speed limit is saturated.  

Including known restrictions on brane inflation and DBI inflation models
will dramatically modify the original phase space estimates of \cite{InitialTuning}, which we trace
back to two effects: First, the total available phase space volume for initial conditions is significantly 
reduced due to effective field theory restrictions \cite{MyersStringyInflation}
as well as compactification effects \cite{BraneGravWaves}.  
Second, the existence of the DBI inflationary attractor in regions where the 
usual slow roll attractor does not exist extends the volume of initial conditions phase space which gives
rise to a sufficient amount of inflation, even when the majority of inflation is obtained in the slow roll regime.

We will also show that ``overshoot trajectories", 
trajectories which overshoot the standard slow roll inflationary region because they are moving
too fast, can in many cases be excluded due both to the restriction on the available phase space
from backreaction effects
and from the presence of the DBI inflationary attractor solutions.

Throughout this paper we will 
develop our formalism and arguments in general, but to better illustrate
our points with a specific example we will examine in detail a potential
with an inflection point.  This is an interesting example to study, because
a.) Inflection point potentials appear in the most
concrete brane inflation models yet constructed \cite{DelicateUniverse,Chasing} (inflection
point potentials can also occur in closed string inflation models, in which inflation is
favored for other reasons \cite{Accidental}), and b.) Inflection point potentials
have many interesting properties, such as overshoot trajectories, to
which we would like to obtain a better understanding by applying the attractor formalism.
It is straightforward to apply our analysis to different choices of potentials and warp
factors.

Note that we are not concerned with making restrictions on phase space or microphysical parameter
space which lead to inflationary models that are consistent with the most current data (see \cite{Numerics} for
recent work in this direction).  We
are only interested in regions of phase space which give rise to a sufficient number
of e-foldings ($60$ or more) and are consistent with known phase space restrictions (such
as Planck-scale effects).  Also note that we are not considering the effect of inhomogenous small scale
and large scale initial conditions or stochastic fluctuations of the
inflaton field (see \cite{Coule} for some initial thoughts in this
area for DBI kinetic terms).  Including these effects is clearly important for making statements about
the generic nature of inflationary trajectories and solutions, and we plan to revisit this
in future work.

The paper is structured as follows.  
In Section \ref{sec:SRAttractor} we review the construction of the inflationary attractor solution
for a scalar field with a canonical kinetic term.  In Section \ref{sec:DBIReview} we review
the DBI inflationary scenario, and construct the DBI inflationary attractor trajectories.  In Section
\ref{sec:PhaseSpace} we show how backreaction effects in string theory reduce
the available volume of initial 
conditions in phase space by several orders of magnitude, thus a significantly increased fraction of phase space
leads to successful brane inflation.  In Section \ref{sec:Overshoot} we show how overshoot trajectories are
excluded in many brane inflation models.  
Finally,
we will conclude with an outlook on future studies in these directions.  

\section{Review of Canonical Kinetic Term Inflationary Attractor}
\label{sec:SRAttractor}
\setcounter{equation}{0}

We begin by reviewing the construction of the slow roll attractor solution discussed in \cite{SRAttractor}.

Standard slow roll inflation is described by the action of a scalar field with canonical kinetic term,
\begin{equation}
S = \int d^4x\ a^3(t) \left(\frac{1}{2}\dot{\phi}^2 - V(\phi)\right)\, ,
\label{eq:SRAction}
\end{equation}
in an expanding background,
\begin{equation}
ds^2 = -dt^2 + a(t)^2 d\vec{x}^2\, .
\label{eq:FRW}
\end{equation}
The equations of motion are
\begin{eqnarray}
&& \ddot{\phi} + 3 H \dot{\phi} + \partial_\phi V(\phi) = 0 \\
&& H^2 = \frac{1}{3 M_p^2} \left(V(\phi) + \frac{1}{2}\dot{\phi}^2\right)\, ,
\end{eqnarray}
and can be written in the Hamilton-Jacobi formalism \cite{HamJac1,HamJac2} in which we exchange $\phi$ for our monotonic time variable
and search for solutions $H(\phi)$ to the first order equations\footnote{Note that not all solutions can be found 
using a Hamilton-Jacobi, or ``pseudo-BPS", formalism;
some specific examples where this formalism fails for multifield scaling cosmologies were found in \cite{HamiltonJac}.},
\begin{eqnarray}
\dot{\phi} &=& - 2 M_p^2 H'(\phi) \\
V(\phi) &=& 3 M_p^2 H^2(\phi) - 2 M_p^4 H'(\phi)^2\, .
\end{eqnarray}

\begin{figure}[t]
\begin{center}
\includegraphics[scale=.4]{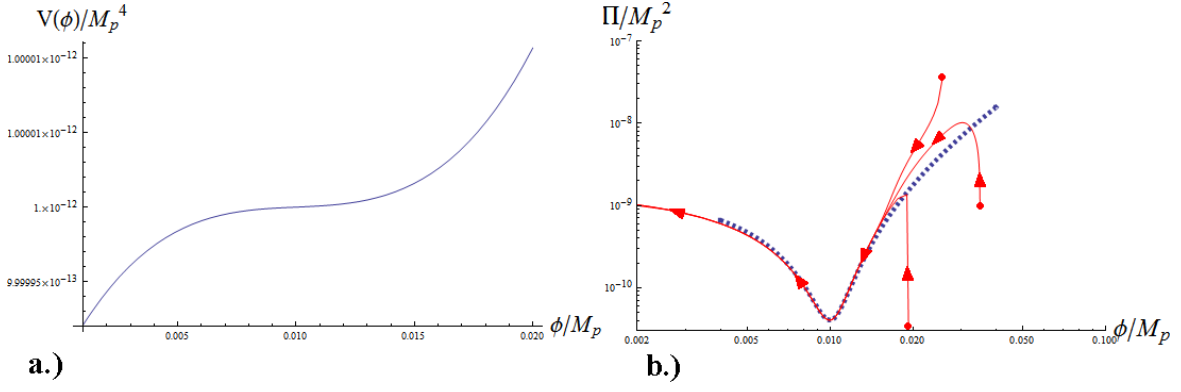}
\caption{\small a.) An inflection point potential $V(\phi) = V_0(1+\lambda_1 (\phi-\phi_0) + \frac{1}{6} \lambda_2 (\phi-\phi_0)^3)$,
with $V_0 = 10^{-12}$, $(\lambda_1,\lambda_2)=(7\times 10^{-5},60)$, $\phi_0 = 0.01$ in Planck units. b.) The phase space diagram for the
potential in a.) for a few sample initial condition choices (solid red).  The inflationary attractor solution is shown
in dashed (blue).  The attractor solution only exists in a finite region corresponding to when the potential allows 
slow roll inflation.  Notice that the solutions flow towards the inflationary attractor solution.}
\label{fig:CanonInflectionAttractor}
\end{center}
\end{figure}

The equations of motion
can also be written as a set of first order differential equations (we are taking a flat universe $k=0$) \cite{SRAttractor}
\begin{eqnarray}
\frac{d\phi}{dt} &=& \Pi\\
\label{eq:CanonAttractEq}
\frac{d\Pi}{dt} &=&-(3H\Pi +V'(\phi))  \\
H^2 &=& \frac{1}{3 M_p^2} \left(V(\phi) + \frac{1}{2}\Pi^2\right)\, .
\label{eq:SRPhaseEqns}
\end{eqnarray}
The solutions to these equations are trajectories in the $(\phi,\Pi)$ phase space, see the example given
in Figure \ref{fig:CanonInflectionAttractor}.

Solutions for which $d\Pi/dt = -(3 H \Pi + V'(\phi)) \approx 0$ are called {\it inflationary attractor
solutions}, and are defined by the trajectories
\begin{equation}
\Pi_{attract} = -\frac{V'(\phi)}{3H}\, ,
\end{equation}
 since small perturbations about this solution $\Pi = \Pi_{attract} + \delta \Pi$ are
driven to zero by the fluctuated form of (\ref{eq:CanonAttractEq}), $d\delta\Pi/dt = - 3 H \delta\Pi$.
It is easy to check that these attractor solutions only exist when the slow roll inflationary conditions
\begin{eqnarray}
\label{eq:SRParam1}
\epsilon_{SR} &=& \frac{M_p^2}{2} \left(\frac{V'(\phi)}{V(\phi)}\right)^2 \ll 1 \\
\eta_{SR} &=& M_p^2 \frac{V''(\phi)}{V(\phi)} \ll 1
\label{eq:SRParam2}
\end{eqnarray}
are satisfied.  In particular, note that
$\Pi_{attract} \sim H^2 \sqrt{\epsilon}$, so the condition for the attractor to be valid
$d\Pi/dt \approx 0$ is satisfied if the slow roll conditions (\ref{eq:SRParam1}-\ref{eq:SRParam2}) 
are also satisfied.

Understanding the behavior of these trajectories can help in understanding the dynamics of the inflationary system and
the fine tuning of initial conditions needed.

\section{D-brane Inflation}
\label{sec:DBIReview}
\setcounter{equation}{0}

We now consider the construction of an action for scalar fields which describe the position of a spacetime-filling
D-brane.

The action for the low energy fields of a D(3+n)-brane (e.g. the scalars parameterizing the position of the brane in the compact
space, where we are ignoring the gauge field on the brane for simplicity) are given by the DBI action,
\begin{equation}
S_{DBI} = - T_{3+n} \int d^{4+n}\xi e^{-\Phi} \sqrt{-\mbox{det}(G_{AB}-B_{AB})}
\end{equation}
where $T_{3+n} = (2\pi)^{-(3+n)} g_s^{-1} \alpha'^{-(2+n/2)}$ is the D-brane tension, $\Phi$ is the dilaton, and
$G_{AB},\ B_{AB}$ are the pullbacks of the metric and the NSNS 2-form onto the brane.  We will consider a GKP-type \cite{GKP}
background
metric of the form of a warped product of a 4-d FRW spacetime and a 6-d space which locally is a cone
over a compact 5-d space\footnote{We will not specify the form of $X_5$ in this paper, but the reader should have
in mind a space like $Y^{p,q}$ or $L^{a,b,c}$.} $X_5$,
\begin{eqnarray}
ds^2 &=& e^{2A(y)} g_{\mu\nu} dx^\mu dx^\nu + e^{-2A(y)} \tilde{g}_{mn} dy^m dy^n \\
\tilde{g}_{mn}dy^mdy^n &=& dr^2 + r^2 ds_{X_5}^2 = dr^2 + \tilde{g}_{ab}dy^a dy^b
\end{eqnarray}
For simplicity we will only consider motion of the D-brane in the radial direction of the throat (for studies which take into
account the angular directions see \cite{D3Vacua,AngularDBI,Spinflation,Pajer}), and will consider only radial
profiles for the warp factor $e^{4A(r)}$ and the dilaton $\Phi(r)$.  We will also take the NSNS 2-form to only have
legs along the angular directions,
\begin{equation}
B_2 = \frac{1}{2} b_{ab} dy^a dy^b
\end{equation}

The DBI action for spacetime-filling $3+n$-branes, with this background, then becomes,
\begin{eqnarray}
S_{DBI} &=& - T_{3+n} \int d^4\xi e^{4A(r)} e^{-\Phi(r)} \sqrt{1+e^{-4A(r)}\partial_\mu r\partial^\mu r} 
	\int d^n \xi \sqrt{\mbox{det}(G_{k\ell} - B_{k\ell})} \nonumber \\
		&=& - T_{3+n} \int d^4\xi e^{4A(r)} e^{-\Phi(r)}B(r) \sqrt{1+e^{-4A(r)}\partial_\mu r\partial^\mu r}
\end{eqnarray}
where
\begin{eqnarray}
G_{k\ell} &=& e^{-2A(r)} r^2 \tilde{g}_{ab} \frac{\partial y^a}{\partial\xi^k}\frac{\partial y^b}{\partial\xi^\ell} \\
B_{k\ell} &=& b_{ab} \frac{\partial y^a}{\partial\xi^k}\frac{\partial y^b}{\partial\xi^\ell} \\
B(r) &=& \int d^n \xi \sqrt{\mbox{det}(G_{k\ell} - B_{k\ell})} \, .
\end{eqnarray}
Notice that for a standard $D3$-brane, $B(r) = 1$.
We can write the action in a more standard form by identifying a scalar field (which is in general non-locally related
to the original radial coordinate),
\begin{eqnarray}
d\phi &=& T_{3+n}^{1/2} B^{1/2}(r) e^{\Phi(r)/2} dr \\
f(\phi) &=& \frac{e^{-4A(r(\phi))}}{e^{-\Phi}B(r(\phi)) T_{3+n}}\, .
\end{eqnarray}
With these definitions, the general DBI action for the radial motion of D-branes which fill the non-compact space is,
\begin{equation}
S_{DBI} = - \int d^4\xi \sqrt{-g_4} f^{-1}(r) \sqrt{1+f(r) \partial_\mu \phi \partial^\mu \phi}
\end{equation}

The interactions of the brane with other background fields (such as the RR 4-form) and (possibly non-perturbative) sources 
leads to the introduction of a potential \cite{DelicateUniverse,Uplifting},
\begin{eqnarray}
S_{DBI} = -\int d^4 x \sqrt{-g_4}\, (f(\phi)^{-1} [\sqrt{1+f(\phi)\partial_{\mu}\phi\partial^{\mu}\phi}-1] + V(\phi)) \, .
\label{eq:Action}
\end{eqnarray}
Note that to lowest order in $f(\phi) \partial_{\mu}\phi\partial^{\mu}\phi \ll 1$, we obtain
the usual canonically normalized kinetic term for the field $\phi$, plus a series of higher dimensional terms.

This action defines a speed limit for the homogenous scalar field $\phi(t)$ \cite{Silverstein},
\begin{eqnarray}
\label{eq:SpeedLimit}
\dot{\phi} &\leq& \frac{1}{\sqrt{f(\phi)}} \\
\gamma &\equiv& \frac{1}{\sqrt{1-f(\phi)\dot{\phi}^2}}\\
\label{eq:Lorentz}
c_s &\equiv& \sqrt{1-f(\phi)\dot{\phi}^2}
\label{eq:soundspeed}
\end{eqnarray}
where $\gamma = \gamma(\phi)$ is the Lorentz factor for the motion of the brane and $c_s$ is called the sound speed.  
For large $f(\phi)$ the speed limit (\ref{eq:SpeedLimit}) forces
the field to move slowly, which leads to the surprising result that with a large enough $f(\phi)$ the field will roll slowly even for a steep
potential $V(\phi)$.  

The equation of motion for a homogenous $\phi$, together with the Friedmann equation, is
\begin{eqnarray}
\label{eq:DBIEOM}
\ddot{\phi}&+&3H\dot{\phi} - \frac{\dot{c}_s}{c_s}\dot{\phi} + c_s \partial_{\phi}\left(V+\frac{c_s-1}{f}\right) = 0\, , \\
H^2 &=& \frac{1}{3 M_p^2}\left(V(\phi) + \frac{\gamma-1}{f(\phi)}\right)
\label{eq:DBIFriedmann}
\end{eqnarray}
The energy density and pressure of the system are 
\begin{eqnarray}
\rho &=& \frac{\gamma}{f}+(V-f^{-1}) \\
p &=& -\frac{1}{f \gamma} - (V -f^{-1})\, .
\end{eqnarray}
One can solve these equations in many cases of interest
using the Hamilton-Jacobi formalism in which we treat $\phi$ as our monotonic time variable and search for solutions $H=H(\phi)$
to the following first order equations \cite{Silverstein},
\begin{eqnarray}
f^{1/2}(\phi) \dot{\phi} &=& -\left(\sqrt{1+\frac{1}{4 M_p^4 f H'^2}}\right)^{-1} \nonumber \\
V &=& 3 M_p^2 H^2 -\frac{2 H'}{f(\phi)} \sqrt{1+\frac{1}{4 M_p^4 f H'^2}} + \frac{1}{f}\, .
\label{eq:HJ}
\end{eqnarray}
This leads to a simple expression for the relativistic gamma-factor of the brane,
\begin{equation}
\gamma(\phi) = \sqrt{1+4 M_p^4 f H'^2}\, ,
\end{equation}
from which we learn that in order to have the speed limiting effect e.g. $\gamma \gg 1$ we need both 
$f(\phi)$ and $H'$ to be large (in Planck units).

\subsection{DBI Attractor Solutions}

Inflationary solutions to the system of equations (\ref{eq:HJ}) are attractor solutions \cite{Meng}; we
repeat the argument of \cite{Meng} here for completeness.
Suppose that $H_0(\phi)$ is a
solution of (\ref{eq:HJ}) and consider a small perturbation about this solution $H(\phi) = H_0(\phi)+\delta H(\phi)$.  The linearized equation
for $\delta H$ becomes,
\begin{equation}
\frac{\delta H'}{\delta H} = \frac{3H_0}{2H_0'}\sqrt{1/M_p^4 + 4 f H_0'^2}
\end{equation}
with the general solution
\begin{equation}
\delta H(\phi) = \delta H(\phi_i)\ \mbox{exp}\left[\int_{\phi_i}^{\phi} 3H_0 \sqrt{1/M_p^4 + 4 f H_0'^2}\frac{d\phi}{2 H_0'}\right]\, ,
\end{equation}
which is just a specific case of the more general result $\delta H\sim e^{-3N_e}$ \cite{HamJac2}\footnote{We would like
to thank the referee from bringing this to our attention.}.
Since the number of e-folds grows as $N_e\sim H_0 t$ during inflation, we see that perturbations 
$\delta H$ are exponentially damped, as expected for attractor solutions.

It is often useful to also know the form that the inflationary attractor takes in field space.
We will consider the first order equations of motion for
the variable $\chi \equiv \sqrt{f} |\dot{\phi}| = \pm \sqrt{f} \dot{\phi}$,
where the plus sign is for
models where the brane starts in the IR and moves towards the UV (large $r$) direction of the throat \cite{IRDBI} 
and the minus sign
is for models \cite{Silverstein} where the brane starts in the UV and moves towards the IR (small $r$) direction.
Note that while $\chi$ is not the canonical momentum associated with the $\phi$ field (because of the non-canonical
form of the equation of motion), it is nevertheless useful for finding attractor solutions to the equations of
motion (\ref{eq:DBIEOM}-\ref{eq:DBIFriedmann}).
The speed limit (\ref{eq:SpeedLimit}) and Lorentz factor (\ref{eq:Lorentz}), rewritten in these coordinates, are
\begin{eqnarray}
\chi &\leq& 1 \\
\gamma &= & \frac{1}{\sqrt{1-\chi^2}}.
\end{eqnarray}
The first order equations of motion in this variable are
\begin{eqnarray}
\chi &\equiv & -\dot{\phi} \sqrt{f}  \\
\frac{d\chi}{dt} &=& (1-\chi^2)\left[V'f^{1/2}(1-\chi^2)^{1/2}-3H\chi - \frac{f'}{f^{3/2}}\left(1-(1-\chi^2)^{1/2}\right)\right], \\
H^2 &=& \frac{1}{3 M_p^2}\left(V(\phi) + \frac{1}{f}(\frac{1}{\sqrt{1-\chi^2}}-1)\right)\, ,
\label{eq:DBIChi}
\end{eqnarray}
where we used the energy conservation equation $\dot{\rho} + 3H(\rho+p) = 0$ for the energy density and pressure.

We will consider ``fixed point" solutions of the type $d\chi/dt \approx 0$, which has a trivial solution $\chi=1$ as well
as a non-trivial solution
\begin{equation}
\chi_{Attract}(\phi) = \frac{(A+\Delta)\sqrt{1+A^2+2A\Delta}-\Delta}{(1+(A+\Delta)^2)}\, ,
\label{eq:chiAttract}
\end{equation}
where we have defined two dimensionless parameters which determine the nature of the solution,
\begin{eqnarray}
A(\phi) &\equiv& \frac{V'(\phi) f(\phi)^{1/2}}{3H(\phi)} \\
\Delta(\phi) &\equiv & \frac{f'(\phi)}{3H(\phi)f^{3/2}(\phi)}\, .
\end{eqnarray}
These paramters are related to the usual DBI inflationary parameters \cite{Shandera}
(which reduce to the slow roll parameters (\ref{eq:SRParam1}-\ref{eq:SRParam2}) when
the motion is non-relativistic)
\begin{eqnarray}
\label{eq:DBIParam1}
\epsilon_D &\equiv & \frac{2 M_p^2}{\gamma} \left(\frac{H'(\phi)}{H(\phi)}\right)^2 \\
\eta_D &\equiv & \frac{2 M_p^2}{\gamma} \left(\frac{H''(\phi)}{H(\phi)}\right) \\
\kappa_D &\equiv & \frac{2 M_p^2}{\gamma} \left(\frac{H'(\phi)}{H(\phi)}\frac{\gamma'}{\gamma}\right)
\label{eq:DBIParam3}
\end{eqnarray}
in the following way (to leading order)
\begin{eqnarray}
A(\phi) &\approx & \epsilon_D^{1/2}\ \sqrt{\frac{2}{3} F(\phi)} \\
\Delta(\phi) &\approx & \frac{\kappa_D}{\epsilon_D} \frac{1}{2 M_p^2 f^{1/2}(\phi) F(\phi)} - 
	\frac{\eta_D}{\sqrt{\epsilon_D}} \frac{1}{\sqrt{6}} \frac{\gamma(\phi)}{\sqrt{F(\phi)}}
\end{eqnarray}
where we defined $F(\phi) \equiv f(\phi) \gamma(\phi) V(\phi)$ for simplicity.

That this is (locally) an attractor solution can easily be seen 
by expanding (\ref{eq:DBIChi}) to linear order in $\delta\chi = (\chi-\chi_{Attract})$,
\begin{eqnarray}
\dot{(\delta\chi)} &\approx & - 3H\lambda \delta\chi\, ,
\label{eq:AttractStability}
\end{eqnarray}
where $\lambda > 0$.
Clearly we have attractive behavior since for $\delta\chi<0$, $\dot{(\delta\chi )} > 0$ which drives the system up to the attractor curve solution, 
and likewise
for $\delta\chi> 0$, $\dot{(\delta\chi)} < 0$ which drives the system down to the attractor curve solution.  
Thus we see that $\chi_{Attract}$ is an attractor
solution.

In a similar way as with the slow roll case, one can show that the attractor solution only exists when the
DBI inflationary parameters Eqs.(\ref{eq:DBIParam1}-\ref{eq:DBIParam3}) are small, corresponding to inflationary
solutions.
Also, note that the relativistic gamma factor along the attractor solution can be written as
\begin{equation}
\gamma_{Attract} = \frac{A+\Delta}{\chi_{Attract}+\Delta} 
\end{equation}

\begin{figure}[t]
\begin{center}
\includegraphics[scale=.45]{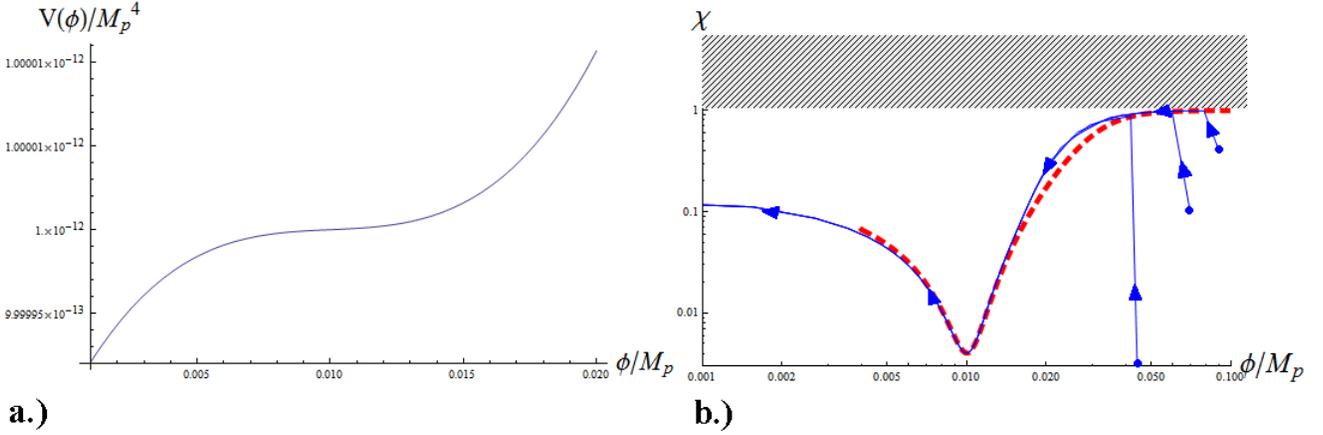}
\caption{\small a.) The inflection point potential of (\ref{eq:InflectPot}).
b.) The pseudo-phase space diagram for a
DBI kinetic term for the
potential in a.) for a few sample initial condition choices (solid blue).  The inflationary attractor solution is shown
in dashed (red).  Notice that the solutions flow towards the inflationary attractor solution, which
interpolates between the DBI and slow roll regimes.  The crossed out region
is inaccessible due to the DBI speed limit, $\chi \leq 1$.}
\label{fig:DBIInflectionAttractor}
\end{center}
\end{figure}

The curve $\chi_{Attract}(\phi)$ traces out a trajectory in the $(\phi,\chi)$ pseudo-phase space
(so called because the coordinate $\chi$ is not the conjugate momentum), 
and depends on the form of the potential and the warp factor through $A(V(\phi),f(\phi))$ and $\Delta(V(\phi),f(\phi))$.
An example of the attractor solution curve which interpolates between the DBI and slow roll regime
is for an inflection point type potential 
(such as may come from a $\bar{D}3$-brane inflating in the background of \cite{DelicateUniverse,Chasing}),
at the tip of a Klebanov-Strassler throat \cite{KS} with a constant warp factor \cite{DBIGap}
where the inflaton is moving in an angular direction as in \cite{Pajer}
\begin{eqnarray}
\label{eq:InflectPot}
V(\phi) &=& V_0(1+\lambda_1 (\phi-\phi_0) + \frac{1}{6} \lambda_2 (\phi-\phi_0)^3) \\
f(\phi) &=& \frac{\lambda}{\mu^4}
\label{eq:WarpFactor}
\end{eqnarray}
with the values of the parameters
$V_0 = 10^{-12}$, $(\lambda_1,\lambda_2)=(7\times 10^{-5},60)$, $\phi_0 = 0.01$ in Planck units
chosen to produce the correct
normalization of density perturbations\footnote{We would
like to thank Daniel Baumann, Hiranya Peiris, and Enrico Pajer for bringing this to our attention.}
and $\lambda = N_{D3} T_3$, $\mu = h_A \lambda^{1/4}$, 
with the effective number of D3-branes of the background 
given as $N_{D3} = 10^2$ and the hierarchy $h_A = 10^{-4}$ ($g_s=10^{-3}, m_s=10^{-0.5}M_p$).
This example is shown in Figure \ref{fig:DBIInflectionAttractor}.
Notice how the attractor solution smoothly connects the DBI regime at large and small $\phi$, where the potential
is steep, with the slow roll regime at intermediate $\phi$, where the potential is flat.

The attractor solution curve (\ref{eq:chiAttract}) is controlled largely by the single parameter $A$, which leads to standard 
slow roll (with canonical kinetic term) when it is small and DBI speed-limiting behavior when it is large:
\begin{eqnarray}
A(\phi) \rightarrow 0&,& \ \chi_{Attract}(\phi) \rightarrow A \rightarrow 0\ \vspace{.1in} \gamma_{attract} \rightarrow 1\ \mbox{Slow Roll} \\
A(\phi)  \gg \Delta, 1&,&\ \chi_{Attract}(\phi) \rightarrow \frac{A}{\sqrt{1+A^2}} \rightarrow 1\ \vspace{.1in} \gamma_{attract} \rightarrow A >> 1\  \mbox{DBI}.
\end{eqnarray}
We see, then, that different limits of $A$ correspond to either slow roll or DBI-like solutions to the equations of motion.

For an energy density dominated by the potential (necessary for accelerated expansion), 
we can write the interpolation parameter $A$ as
\begin{equation}
A = \frac{M_p}{\sqrt{3}}\frac{V'}{V}\sqrt{V f} = \sqrt{\frac{3}{2}\epsilon_{SR}}\sqrt{\frac{V}{T_{D3,local}}}
\end{equation}
where $\epsilon_{SR} \equiv M_p^2/2 (V'/V)^2$ is the standard slow roll parameter which must be much less
than one for slow roll inflation.  We see, then, that DBI behavior emerges when the combination of the standard slow
roll parameter and
the potential in units of the local brane tension is much larger than one (indicating that slow roll is no longer valid).
Notice that for small $A$, the attractor solution
\begin{equation}
\chi_{Attract} = A = \frac{V'f^{1/2}}{3H}
\end{equation}
is just the slow roll attractor equation, as should be expected.

In the limit $\Delta \rightarrow 0$, which is the limit in which the dependence of the warp
factor on the inflaton field becomes negigible, the attractor solution curve takes the form
\begin{equation}
\chi_{Attract}(\phi) = f^{1/2} |\dot{\phi}| \rightarrow \frac{1}{\sqrt{1+1/A^2}} = \left(\sqrt{1+\frac{9 H^2}{f V'^2 }}\right)
\end{equation}
which is the same as the Hamilton-Jacobi solution (\ref{eq:HJ}) in this limit.  Thus we
see that the attractor solutions give the Hamilton-Jacobi solutions.
Also note that the k-inflationary attractor solutions described by \cite{kinflation} are not
the same as the attractor solutions described here in (\ref{eq:chiAttract}), which can be clearly seen 
since our attractor solutions vanish in the limit of vanishing
potential, $A\rightarrow 0$, whereas k-inflationary attractors are present without a potential.
It would be interesting to construct a generalized attractor solution for a general non-canonical lagrangian
which incorporates both of these known solutions.

The number of e-folds along a given trajectory is calculated from
\begin{equation}
N_e = \int H[\phi(t)] dt = \int \frac{H[\phi]}{\dot{\phi}} d\phi = \int \frac{H[\phi]}{\chi(\phi)}\sqrt{f(\phi)} d\phi\, .
\end{equation}
Since the DBI kinetic term restricts $\chi \leq 1$ along the inflationary attractor, while $\chi$ can be
greater than one for the canonical attractor, we see that D-brane inflationary attractors generically
lead to more e-folds than canonical attractors for the same choice of
potential since $\chi$ is much smaller for the former.  

\begin{figure}[t]
\begin{center}
\includegraphics[scale=.45]{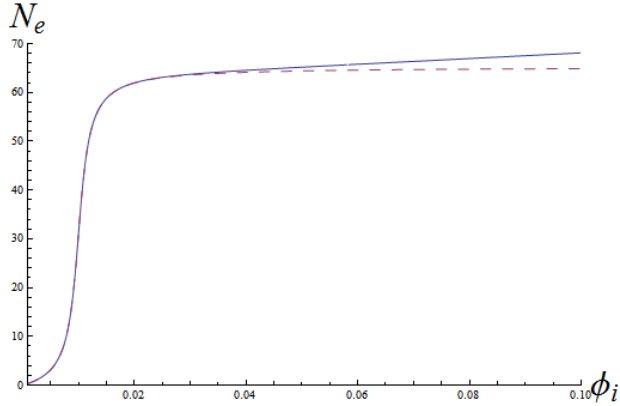}
\caption{\small 
The number of e-folds as a function
of the starting value of the field, $\phi_i$, along the DBI attractor (solid blue) 
is slightly greater than for the canonical attractor (dashed red) because of the speed limit restriction $\chi \leq 1$.
Also note that the majority of the e-folds are generated in the standard slow roll region near the inflection
point, even for the brane inflation model with DBI kinetic term.}
\label{fig:EfoldAttractor}
\end{center}
\end{figure}

\section{Phase Space and Initial Conditions}
\label{sec:PhaseSpace}
\setcounter{equation}{0}

In general, picking a random point in phase space for an initial condition of the inflaton will
not lead to enough inflation.  The initial conditions leading to 60 or more e-folds is typically
a small subset of the full space of possible initial conditions.  This leads to an {\it initial conditions fine-tuning problem}:
for a given model, what amount of fine tuning of (homogenous) initial conditions is needed to give 60 or more e-folds
of inflation?  
In this section, we will revisit this question for brane inflation models.  We will find that the 
initial conditions fine-tuning for the inflaton field is improved by many orders of magnitude by two effects:
\begin{enumerate}
\item Brane inflation, being a UV-complete theory, has stringent effective theory bounds on the available initial
conditions phase space.  In particular, we will see that requirements from compactification place stringent bounds
on the available field space range, and stringy effects place stringent bounds on the available momentum
and field space range.  These effects shrink the overall available volume of initial conditions phase space, increasing
the fraction of phase space which leads to sufficient inflation.
\item The existence of DBI inflationary attractor solutions in regions of phase space where there
is no slow roll inflationary attractor (such as when the potential is steep), increases the
overall total number of e-folds for a given set of initial conditions, increasing the volume of
phase space in which sufficient inflation occurs.
\end{enumerate}

In the remainder of this section, we discuss how these effects appear in more detail.

\subsection{Available Phase Space}

We first will analyze the available phase space for canonical kinetic term 
scalar fields, as well as brane-inflation scalar fields.  In what follows, by
{\it canonical} scalar fields, we mean scalar fields which have the usual kinetic
term $1/2 \dot{\phi}^2$ with no higher derivative corrections, as we would typically
write down in any standard field theory inflation model.  By {\it brane inflation} scalar fields
we mean scalar fields that have a DBI kinetic term (\ref{eq:Action}), regardless of whether
the e-folds are generated in the slow roll or DBI inflationary regimes.  In fact, for the example
shown in Figure \ref{fig:EfoldAttractor} and which will be analyzed in more detail later, 
it is clear that the majority of the e-folds are generated
in the slow roll region.  Nevertheless, we will see that the effects from the DBI kinetic term
still lead to significant improvements of initial conditions fine tuning.  This also
suggests the interesting possibility that DBI inflation may be important for generating the observable
e-folds in the CMB, while slow roll inflation generates the remaining 50-55 e-folds necessary to solve the
flatness and horizon problems (this was also suggested recently in \cite{Pajer}).  This can
have interesting consequences for observation of primordial non-Gaussianity and other unique DBI observables.

For canonical kinetic term scalar fields, it is not clear what restrictions to place on the initial conditions without
knowing the UV completion of the theory.  Nonetheless, we 
will restrict ourselves to small field models in which the field space range is bounded by
$\Delta\phi \leq M_p$.
Furthermore, we generally expect that the Hubble parameter $H$
must not be bigger than the Planck scale, at which point higher dimensional operators involving the curvature will
start to become important.  This imposes a restriction on the amount of intial kinetic energy allowed
for canonical inflaton fields,
\begin{equation}
\frac{1}{2} \Pi_{canon}^2 \leq M_p^4\, .
\end{equation}
The limits on the field-space and momentum for a scalar field with a canonical kinetic term are then
(we consider here only positive momentum for simplicity)
\begin{eqnarray}
0\leq \phi &\leq & M_p \\
0\leq \Pi_{canon} &\leq & M_p^2\, .
\end{eqnarray}

\begin{figure}[t]
\begin{center}
\includegraphics[scale=.5]{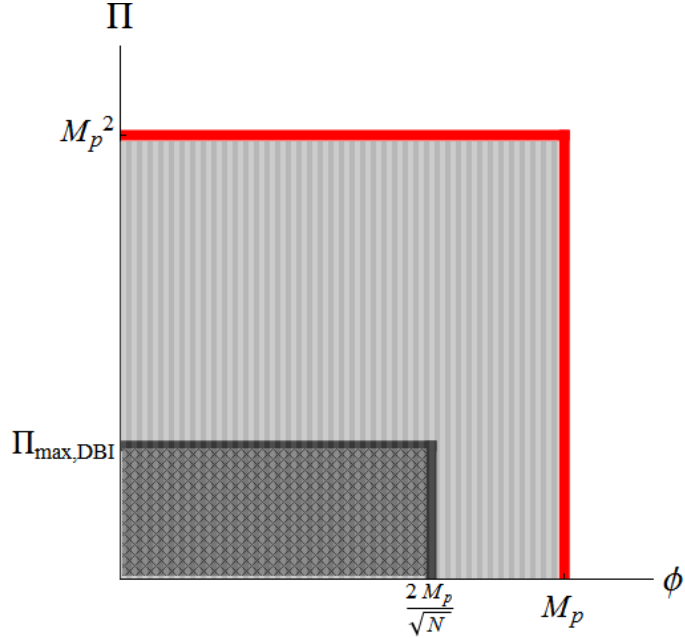}
\caption{\small The available initial conditions phase space is shown for the canonical kinetic term (red outline, vertical
gray lined area) and 
brane-inflation/DBI kinetic term (black outline, gray cross-hatched area).  Notice that the available phase space allowed for brane inflation
is much smaller in volume than for the standard field theoretic model due to effective field theory, compactification
and stringy effects, by a factor of $2 \Pi_{max,brane}/\sqrt{N_{D3}} \ll 1$.}
\label{fig:AvailPhaseSpace}
\end{center}
\end{figure}

For the simplest brane inflation models, compactification constraints typically require $\Delta\phi \leq M_p$ at most, which
comes from the fact that a brane cannot move a larger distance than the size of the manifold
in which it is embedded.
However, the most well understood brane inflation models are typically constructed from D3-branes 
in a warped throat, where a more restrictive bound on the field space range $\Delta \phi \leq 2 M_p/\sqrt{N_{D3}}$
\cite{BraneGravWaves} exists, where $N_{D3}$ is the effective D3-brane charge
of the warped background\footnote{Wrapped D-brane models, however, which modify the normalization of the canonically
defined scalar field by factors relating to the volume of the wrapped space (with fluxes), may be able to evade
this constraint (although concerns with backreaction are an issue; see \cite{WrappedDBI} for recent work).}.  
Since $N_{D3} \gg 1$ is necessary in order to trust the supergravity description of
the warped throat, this bound is significantly more restrictive.
We will take $\phi \leq 2 M_p/\sqrt{N_{D3}}$ as a restriction on our available field space since this corresponds
to the most well-developed D3-brane models, but it is easy to allow for other models by removing the
$2/\sqrt{N_{D3}}$ factors which show up in the phase space volume estimates.

The conjugate momentum to $\phi$ for D-brane inflation models is
\begin{equation}
\Pi_{brane} \equiv \frac{\dot{\phi}}{\sqrt{1-f(\phi)\dot{\phi}^2}} = \frac{1}{\sqrt{f(\phi)}}\frac{\chi}{\sqrt{1-\chi^2}}\, ,
\end{equation}
which reduces in the slow roll limit $\Pi\approx \dot{\phi}$ to the usual scalar field momentum.
D-brane inflation models have stringent effective theory constraints on the allowed Hubble rate - in particular,
we must require that the Hubble rate is smaller than the local (warped) string scale or Kaluza-Klein (KK) scale so
that we do not produce (warped) strings or KK modes from the expanding
background which can backreact on the geometry and cast doubt on the 4-dimensional
description \cite{MyersStringyInflation, WrappedDBI}.  Thus, the restriction for not producing
warped strings in the throat we are using $H\leq m_{s,local} = m_s h_A$ leads to a restriction
on the available momentum accessible to the brane inflaton field, $\Pi \leq \Pi_{max,brane} = M_p m_{s,local}$.
Our restrictions then on the allowable phase space for brane inflaton fields become
\begin{eqnarray}
0\leq \phi &\leq & \frac{2 M_p}{\sqrt{N_{D3}}} \\
0\leq \Pi_{brane} &\leq & \Pi_{max,brane} = M_p m_{s,local}\, .
\end{eqnarray}

Since $m_{s,local}\ll M_p$, the ratio of the areas of the available brane inflation phase space
to that of the canonical available phase space in the simplest scenario of
a D3-brane is 
\begin{equation}
\frac{V_{phase,brane}}{V_{phase,canon}} \sim \frac{2}{\sqrt{N_{D3}}}\frac{\Pi_{max,brane}}{M_p^2} \sim \frac{2}{\sqrt{N_{D3}}} \frac{m_{s,local}}{M_p} \ll 1\, ,
\end{equation}
which decreases the available phase space by several orders of magnitude, a significant effect.
These restrictions on the available initial conditions phase space are shown in Figure \ref{fig:AvailPhaseSpace}.
Decreasing the total available phase space will lead to a corresponding improvement in the amount of required fine tuning
of initial conditions, as we will see in more detail in the next subsection.

\subsection{Initial Conditions Fine Tuning}

In the previous section, we noticed that brane inflation models have stringent
effective theory constraints which severely restrict the available phase space for initial
conditions.
In most cases, this
is the dominant effect which increases the percentage of initial conditions phase space leading
to (enough) inflation, thus decreasing the overall amount of fine-tuning needed.

The increase in the total number of e-folds along the D-brane inflationary trajectory also can play a role
in reducing the amount of fine tuning.
Since the DBI kinetic term is inflationary for a larger
region of phase space, solutions flow to an inflationary attractor solution sooner than in the canonical
case, increasing the amount of phase space that leads to inflationary solutions and increasing the total
number of e-folds in any given region of phase space.

\begin{figure}[t]
\begin{center}
\includegraphics[scale=.45]{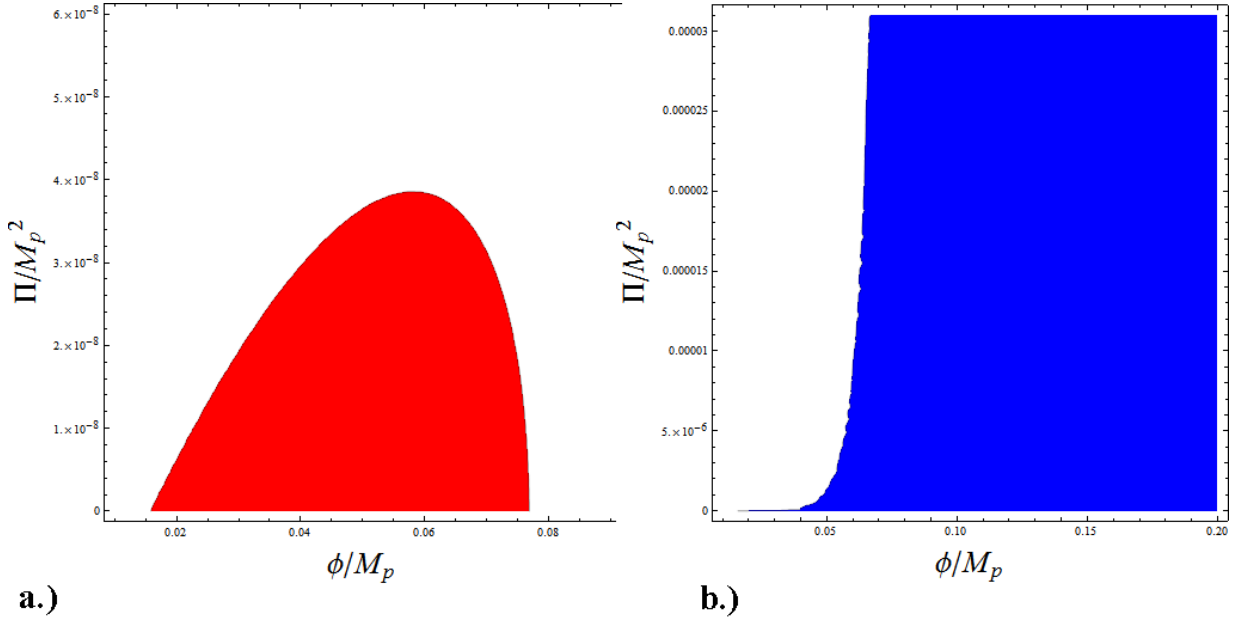}
\caption{\small The shaded region shown is the region of phase space that leads to 60 e-folds of inflation for the case
of a scalar field with a canonical kinetic term ((a), red) and for brane inflation with a DBI kinetic term ((b), blue) for
the same choice of potential and warp factor as described in (\ref{eq:PhaseSpacePotential}-\ref{eq:PhaseSpaceWarp}).
The region which leads to 60 e-folds of inflation is a significantly larger
fraction of the total for the D-brane scalar field, implying that the initial conditions fine tuning for
the brane inflation model is significantly reduced.}
\label{fig:PhaseSpaceFineTuning}
\end{center}
\end{figure}

We will quantify the amount of phase space fine tuning necessary to get sufficient inflation (defined to
be $60$ e-folds) for our given models.  The quantity of interest will be the fraction (or percent) of phase space
that leads to sufficient inflation, e.g.\footnote{This quantity $f_X$, a pure number, should not be confused with the ``warp factor"
$f(\phi)$ which is a function of $\phi$.  We have endeavored to make the functional form of the warp factor explicit throughout
to remove this potential confusion.}
\begin{equation}
f_X = \frac{\mbox{Volume of phase space for system X giving sufficient inflation}}{\mbox{Total volume of phase space for system X}}\times 100\%\, .
\label{eq:FineTune}
\end{equation}
Since we wish to make minimal assumptions about the pre-inflationary era, this quantity treats all initial conditions
as equally likely.  Specific assumptions, such as imagining a cold pre-inflationary era or setting initial conditions with the
wavefunction of the Universe, will place different measures on the likelihood of inflationary initial conditions, and we will
not explore these possibilities here.  
As we will see, including the brane inflation effects discussed above will largely reduce
the need for such assumptions in improving (homogenous) initial conditions fine tuning.
There is considerable ambiguity about the right measure to use for quantifying inflationary fine tuning (for a recent discussion
of some of the problems see \cite{Measure}) and we do not claim
that the simple measure defined above is correct.  Nevertheless, it serves as a useful diagnostic for comparing some
level of ``fine tuning" between the canonical and brane inflation systems.

To compare the fine tuning between canonical kinetic term and D-brane systems we will use the {\it same potential} for each system, and
compare the fraction above.  As an illustrative example we will take the same potential and warp factor as before,
\begin{eqnarray}
\label{eq:PhaseSpacePotential}
V(\phi) &=& V_0(1+\lambda_1 (\phi-\phi_0) + \frac{1}{6} \lambda_2 (\phi-\phi_0)^3) \\
f(\phi) &=& \frac{\lambda}{\mu^4}
\label{eq:PhaseSpaceWarp}
\end{eqnarray}
$V_0 = 10^{-12}$, $(\lambda_1,\lambda_2)=(7\times 10^{-5},60)$, $\phi_0 = 0.01$ in Planck units
and $\lambda = N_{D3} T_3$, $\mu = h_A \lambda^{1/4}$, 
with the effective number of D3-branes of the background 
given as $N_{D3} = 10^2$ and the hierarchy $h_A = 10^{-4}$ ($g_s=10^{-3}, m_s=10^{-0.5}M_p$).
The number of e-folds for a given set of initial conditions can be numerically evaluated;
those initial conditions which give rise to 60 or more e-folds are shown in Figure
\ref{fig:PhaseSpaceFineTuning}.  Similar diagrams can easily be constructed for other choices
of the warp factor and potential.

We can estimate the
required tuning as follows: let us overestimate the region which gives rise to sufficient inflation for the
canonical case as being in the rectangular region (all numerical values
are given in Planck units) $0.015\leq \phi \leq 0.075$ and $0\leq \Pi \leq 4\times 10^{-8}$.  
The overall
available phase space is the rectangular region $0\leq \phi \leq 1$ and $0\leq \Pi\leq 1$,
The fraction
of phase space that gives sufficient inflation for the canonical case for this example is then no larger than
\begin{equation}
\label{eq:CanonTuning}
f_{canon} \approx \frac{0.06\times (4\times 10^{-8})}{1}\times 100\% \approx 2.4\times 10^{-7}\%\, ,
\end{equation}
which clearly is a system that requires fine tuning of initial conditions.

For the D-brane system, we see from Figure \ref{fig:PhaseSpaceFineTuning} that the phase space that gives sufficient
inflation is roughly the rectangular region
$0.05\leq \phi \leq 0.2$ and $0\leq \Pi \leq 3.1\times 10^{-5}$, while the total available phase space is the region
$0\leq \phi \leq 0.2$ and $0\leq \Pi \leq m_{s,local}\approx 3.1\times 10^{-5}$.  
The fraction of phase space that gives sufficient inflation
for the DBI case is then approximately,
\begin{equation}
f_{brane} \approx \frac{0.15\times (3.1\times 10^{-5})}{.2\times (3.1\times 10^{-5})} \times 100\% \approx 75\% \, .
\end{equation}
Thus, we see that this DBI/brane inflation model is essentially entirely free of (homogenous) 
initial conditions fine tuning.
We also see that the DBI kinetic term improved the fine tuning problem by several orders of magnitude.

More generally, the improvement on the amount of fine tuning required is given by the expression,
\begin{eqnarray}
\frac{f_{brane}}{f_{canon}} 
	&\approx & \frac{\sqrt{N_{D3}}}{2} \frac{M_p^2}{\Pi_{max,canon}} \left(\frac{\mbox{Brane phase space volume for $60+$ E-folds}}{\mbox{Canonical phase space volume for $60+$ E-folds}}\right)\, .
\label{eq:FineTuning}
\end{eqnarray}
The terms not in parentheses come from the reduction in the overall available volume of phase space,
and is mostly independent of the specific model and potential that we choose.  The second term is model- and potential-dependent
so it is harder to estimate in general, but it seems reasonable that the volume of phase space that gives rise to 
60 or more e-folds of inflation for
a canonical inflaton system should be at least the same size as the corresponding volume for brane inflation
models since the attractors are identical in the small $\Pi$ region, but may be somewhat bigger.
Estimating
$$
\frac{\mbox{Canonical phase space volume for 60+ E-folds}}{\mbox{Brane phase space volume for 60+ E-folds}}\sim {\mathcal O}(10^{1})
$$ 
at most and $\Pi_{max,canon}/M_p^2 = h_A m_s/M_p\sim {\mathcal O}(10^{-2}-10^{-3})$, $N_{D3}\sim 10^2$
we find that it is reasonable for the the improvement in the initial condition fine tuning to be of the order
$$
\frac{f_{brane}}{f_{canon}} \approx {\mathcal O}(10^{2-3})\, ,
$$
which clearly underestimates the improvement in the specific example given above; 
thus initial conditions fine tuning in brane inflation models is better than ``canonical" field theory (small field) 
inflation models by several orders of magnitude.

\subsection{Possible Resolution to the Overshoot Problem}
\label{sec:Overshoot}

\begin{figure}[t]
\begin{center}
\includegraphics[scale=.5]{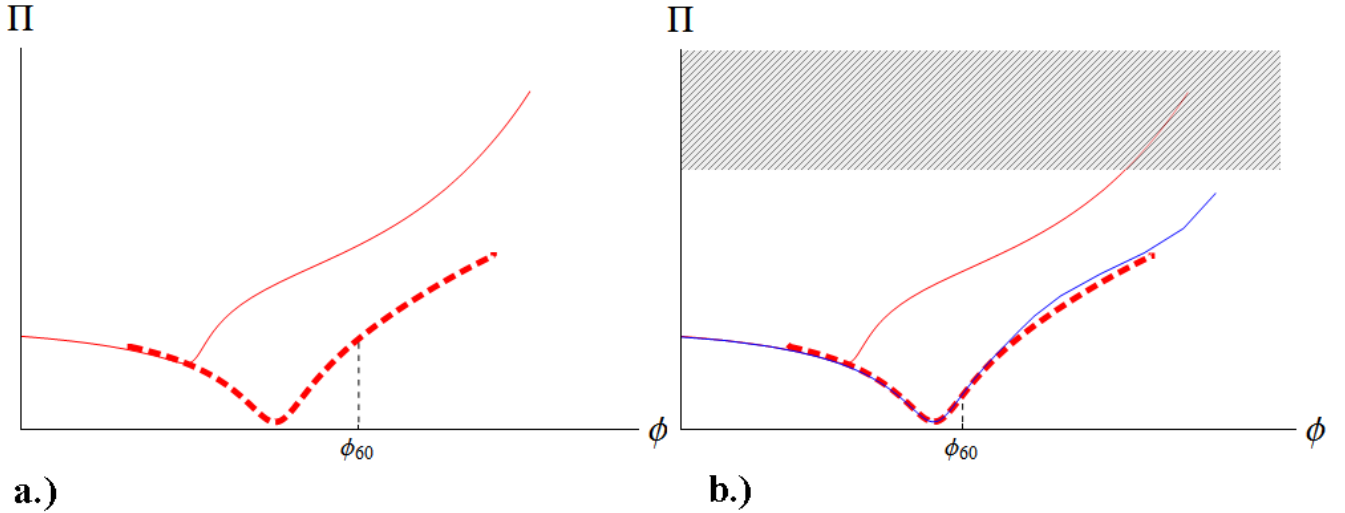}
\caption{\small a.) The overshoot problem occurs when the inflaton initial conditions are such
that the solution (solid red) does not reach the inflationary attractor solution (dashed red) until
after the field value where $60$ e-folds can be obtained along the attractor, denoted here by $\phi_{60}$.
b.) Brane inflation generically avoids the overshoot problem because (i) the initial conditions which
lead to overshoot solutions are excluded by backreaction and effective theory constraints, and (ii) more e-folds are
typically generated, thus the critical field value $\phi_{60}$ decreases somewhat.  The dashed (red) line is the DBI attractor
solution, and the solid (blue) line is a solution for some randomly chosen initial condition consistent
with the speed limit and backreaction constraints.  The crossed out area corresponds to the initial conditions inaccessible to
brane inflation due backreaction.}
\label{fig:OvershootCompare}
\end{center}
\end{figure}

Overshoot 
trajectories occur when a given set of initial conditions does not allow the inflaton
field to converge onto the inflationary attractor solution soon enough to accumulate 60 e-folds, 
or even any at all.
Another way of saying this is that there is a maximum value of the inflation field, $\phi_{60}$,
such that if any given solution is on the attractor solution for $\phi \geq \phi_{60}$
then there will sufficient inflation, but if the solution is only on the attractor
for $\phi \leq \phi_{60}$, then there will not be enough inflation.  The overshoot problem
occurs when the initial conditions for the inflaton are such that the solution only reaches
the attractor solution for $\phi \leq \phi_{60}$ (if even at all).  An example of a typical 
overshoot solution
for a canonical kinetic term is shown in Figure \ref{fig:OvershootCompare} for the same inflection 
point-type potential we have been considering (see \cite{Chasing} for a discussion of overshoot trajectories
in brane-inflation inspired inflection point potentials).

The prevalence of overshoot trajectories is quantified in the same manner as the initial conditions fine tuning
discussed in the previous subsection - note that the definition of the overshoot trajectory given
above is essentially the same as the statement that the inflationary system requires fine tuning of its
initial conditions, just from a slightly different perspective.  
Quantification follows from the analysis above (with the same caveats) by measuring
the quantity $f_X$ in Eq.(\ref{eq:FineTune}) for a given system.  Thus systems which require a high amount of initial
conditions fine-tuning are dominated by overshoot trajectories, and visa versa.

Notice in Figure \ref{fig:OvershootCompare} that overshoot trajectories typically occur when the initial
conditions are such that the initial momentum $\Pi$ is large.  Herein we see a possible resolution
to the overshoot problem in brane inflation: because of the limit on the momentum from string theory backreaction
effects,
initial conditions which lead to overshoot solutions are excluded.  Further, since the DBI inflationary
attractor trajectory typically exists over a larger region of phase space than the canonical attractor
(and leads to more e-folds than the canonical case, as discussed in the previous section), the critical
field value $\phi_{60}$ decreases somewhat so overshoot trajectories are harder to find.
Thus, we see that many brane inflation models may relax the difficulties due to the overshoot problem.

\section{Conclusion}
\setcounter{equation}{0}

By constructing explicit D-brane inflationary attractor trajectories in phase space, 
which reduce to the usual slow roll attractor trajectories in the appropriate limit,
and plotting these attractor trajectories in phase space it is easy to see
when any particular model enters the DBI inflationary regime.  We showed the behavior
of the attractor trajectory for a simple inflection point potential, 
which is well-motivated by recent explicit
constructions of D3-brane inflation in warped throats \cite{DelicateUniverse,Chasing,Pajer}.
In particular, we see clearly from 
Figure \ref{fig:DBIInflectionAttractor}
that this system has a DBI regime and a slow roll regime, connected smoothly through the 
DBI inflationary attractor solution.

We also considered restrictions on the available (homogenous) initial conditions phase space 
for brane inflation models due to compactification and stringy effects, and found that these effects
significantly reduce the available phase space for generic models compared to field theory
models of inflation, Figure \ref{fig:AvailPhaseSpace}, by
several orders of magnitude.
We then compared the volume of homogenous initial conditions in phase space which led to significant ($\geq 60$ e-folds)
of inflation for the previously discussed brane inflation scenario and its canonical field theory kinetic term counterpart.
The fraction of initial conditions volume in phase space which gives rise to sufficient inflation
can be significantly larger for brane inflation models, even those dominated by slow roll,
for two reasons: First, compactification and stringy effects
reduce the overall available phase space, and second, the existence of the brane inflationary attractor for the DBI kinetic term
in regions
where the usual slow roll attractor does not exist increases the overall fraction of phase space which
gives rise to sufficient inflation, even if most of the e-folds are generated in the slow roll regime.

We also proposed a possible solution to the overshoot problem of inflection point-type potentials, in
which the inflaton is rolling too fast when it reaches the flat part of the potential to enter the
slow roll regime.  The resolution is due to the two effects discussed above, namely that 
stringy backreaction effects place restrictions on the allowed momentum of the field so that
most overshoot trajectories are excluded, and the existence of a brane inflationary attractor which
is smoothly connected to the slow roll region efficiently guides inflationary trajectories
into the slow roll regime.

Certainly, the study of inflationary initial conditions must be extended to include inhomogenous
initial conditions, since inflation is supposed to solve the homogeneity and isotropy problems,
not be subject to them.  It would be interesting to extend previous analyses of inhomogenous initial
conditions \cite{InhomogenAttractor} to the brane inflation and see if the DBI kinetic term plays a significant
role.  Stability of the inflationary solution to small fluctuations of the field throughout inflation 
is also important, and can play role in determining which types of inflationary trajectories are more
likely; see \cite{StochasticDBI} for recent work in this direction.
It would also be interesting to scan many different types of brane inflation scenarios (particularly
those with realistic and explicit constructions) to see how important and generic the effects we have
pointed out are.
We intend on investigating these issues in the future.

\section*{Acknowledgments}
It is a pleasure to thank
Daniel Baumann,
Steve Kecskemeti,
Louis Leblond,
John Maiden,
Paul McGuirk,
Enrico Pajer,
Hiranya Peiris,
Sarah Shandera,
Gary Shiu,
and Thomas Van Riet
for many interesting discussions and comments.
B.U. is supported in part by NSF CAREER Award No. PHY-0348093, DOE grant DE-FG-02-95ER40896, 
a Research Innovation Award and a Cottrell Scholar Award from Research Corporation.

\end{document}